\documentclass[runningheads]{llncs}
\usepackage{graphicx}
\usepackage{bookmark}
\usepackage{times}
\usepackage{xcolor}
\usepackage{verbatim}
\usepackage[utf8]{inputenc}
\usepackage{hyperref}
\usepackage{amssymb}
\usepackage{listings}
\usepackage{nameref}
\usepackage{svg}
\usepackage{calc}
\usepackage{newfloat}
\usepackage{makecell}
\usepackage{subfig}
\usepackage{multirow}
\usepackage{float}
\usepackage{amssymb}
\usepackage{csquotes}
\restylefloat{table}

\begin{document}
\title{Developer Operations and Engineering Multi-Agent Systems}
\author{Timotheus Kampik\inst{1} \orcidID{0000-0002-6458-2252} \and\\Cleber Jorge Amaral\inst{2,3} \orcidID{0000-0003-3877-6114}\and\\Jomi Fred Hübner\inst{2} \orcidID{0000-0001-9355-822X}}

\institute{
	Umeå University, Sweden
	\and
	Federal Institute of Santa Catarina, Brazil
	\and
	Federal University of Santa Catarina, Brazil
	\\
    \email{tkampik@cs.umu.se, cleber.amaral@ifsc.edu.br, jomi.hubner@ufsc.br}
}
\authorrunning{T. Kampik \emph{et al.}}
\titlerunning{DevOps and EMAS}
\maketitle
\begin{abstract}
In this paper, we propose the integration of approaches to Engineering Multi-Agent Systems (EMAS) with the \emph{Developer Operations} (DevOps) industry best practice.
Whilst DevOps facilitates the organizational autonomy of software teams, as well as the technological automation of testing, deployment, and operations pipelines, EMAS and the agent-oriented programming paradigm help instill autonomy into software artifacts.
We discuss the benefits of integrating DevOps and EMAS, for example by highlighting the need for agent-oriented abstractions for quality assurance and test automation approaches. 
More generally, we introduce an agent-oriented perspective on the DevOps life-cycle and list a range of research challenges that are relevant for the integration of the DevOps and EMAS perspectives.

\keywords{Agent-oriented Programming \and Engineering Multi-Agent Systems \and Developer Operations}
\end{abstract}
\section{Introduction}

On August 1, 2012, the financial technology venture \emph{Knight Capital Group, Inc} executed a malfunctioning update of their autonomous trading system that caused the large-scale issuing of erroneous orders, leading to losses of more than \$450 million within less than one hour~\cite{10.1257/jep.27.2.51}.
In the software engineering community, the root cause of the error is ascribed to problematic software development processes that do not ensure a sufficient degree of quality assurance automation and testing at different development and deployment stages~\cite{Bass2015}.

To address this and similar issues\footnote{We observe that Knight Capital's system is one of numerous autonomous software systems already in operation within socio-technically complex organizations~\cite{DelaPrieta2019}.}, new software development practices have emerged during the the last decade, most notably the Developer Operations (DevOps) approach~\cite{ebert2016devops}.
DevOps aims to reduce the time for deploying high-quality (validated and verified) software artifacts (and their updates) to complex and heterogeneous \emph{production} environments~\cite{Bass2015}.
Desirable qualities of DevOps-oriented software engineering are reliability, predictability and security~\cite{NicoleForsgren2019}.
For example, DevOps facilitates the autonomy of teams and their individual members to prevent, discover, and fix software bugs quickly and effectively~\cite{NicoleForsgren2019}.
However, one may say that even applying the best industrial-scale software engineering processes in combination with traditional programing paradigms cannot fully prevent problems like the one that occurred during the Knight Capital incident.
Indeed, from an artificial intelligence perspective, an alternative root-cause is the \emph{single-mindedness} and lack of meaningful goal-orientation of the software subsystem (or: agent) that kept issuing orders, without re-assessing over time whether doing so is aligned with the overall objectives of the trading system.
From this perspective, it can be questioned whether the application of the current conception of DevOps is sufficient to ensure quality, and to facilitate the fast-paced development of highly autonomous software systems.

Consequently, one may call for the application of approaches to Engineering Multi-Agent Systems (EMAS) that treat the \emph{agency} of autonomous software artifacts, as well as the environments and organizations these artifacts act in, as first-class abstractions.
Along these lines, this paper proposes a bridge between DevOps and EMAS, with the aim to address the need for a robust method for delivering autonomous software artifacts faster and safer.
Nevertheless, this paper attempts to maintain a critical perspective on the mainstream-readiness of EMAS.
Indeed, the lack of industry-scale tools for engineering autonomous software curbs EMAS adoption in practice~\cite{engineering-gsi-article-2019,logan2018agent}, and we argue that the application of EMAS should always consider efforts to mature EMAS tooling as a prerequisite.
%

\section{DevOps}
\label{background-devops}
Developer Operations (DevOps) describes the industry best practices that integrate software development, quality assurance and operations teams, from both organizational and technological perspectives~\cite{ebert2016devops}.
DevOps can be considered a continuation of the trend towards iterative software development, which started at the turn of the century with the publication of the Agile Manifesto~\cite{DINGSOYR20121213}.
In particular because iterative software development approaches require a fast-paced transition between requirement adjustments, software changes, tests, and deployments, handovers across traditional organizational and technological boundaries become increasingly impractical.
To address this issue, DevOps recommends the integration of software developers, Quality Assurance (QA) engineers, and system administrators into autonomous cross-functional teams that are in charge of developing, testing, deploying, and operating a system or system component~\cite{Bolscher2019}.
This stands in contrast to traditional approaches that segment functional specializations and hence require frequent handovers between teams or even departments, all of which are in charge of one specific task~\cite{Pettigrew2000}.
To support cross-functional teams with the broad range of tasks that fall into the DevOps scope, a plethora of tools exists, many of which have found wide-spread adoption.
For example, continuous integration tools and services allow for the configuration of automated tests and deployments using simple declarative specification and script languages~\cite{6802994}, whereas containerization~\cite{merkel2014docker} and container orchestration tools~\cite{10.1145/2806777.2809955} help speed up and automate the deployment and scaling of complex IT systems across heterogeneous infrastructure.

\begin{figure}[ht]
	\centering
	\includegraphics[width=0.5\textwidth]{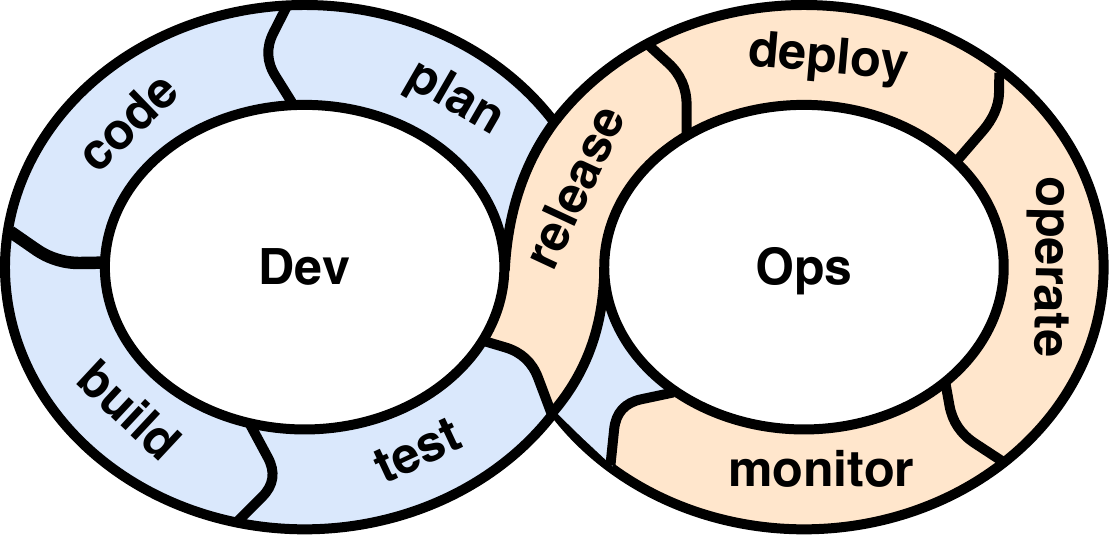}
	\caption{DevOps life-cycle, based on~\cite{ebert2016devops}.}
	\label{fig:devops}
\end{figure}

The DevOps development life-cycle (illustrated in Figure~\ref{fig:devops}) can be described as follows:
\begin{description}
    \item[Plan and code.] DevOps development teams implement features in fast, incremental iterations, which is facilitated by the organizational structure and technological setup. As a consequence, DevOps reduces the overhead of QA, releases, and deployments.
    \item[Build and test.] Each update of the code base triggers the automated execution of one or several test suites.
    Ideally, all technical aspects of software artifact generation (build) and quality assurance are executed automatically; passing tests and builds imply that the software artifact works reliably and can be released without concerns. This requires the development team to treat QA as a key responsibility.
    \item[Release and deploy.] After tests and builds have been successfully executed, deployments (for example to cloud environments) and/or releases (e.g., to package management services) are triggered in an automated or semi-automated manner.
    \item[Operate and monitor.] During operations, a key feature of DevOps is the automation of many system administration tasks, like the provision of additional resources if the load on the system increases. To reduce the overhead of system administration, teams often rely on cloud-based service offerings that abstract away technical details.
\end{description}
Figure~\ref{fig:devops} depicts the DevOps life-cycle.

\section{EMAS}

During the past decades, the EMAS sub-field has emerged as a research direction within the field of artificial intelligence~\cite{shehory2016agent}.
One of the key lines of work within EMAS is the refinement of the Agent-Oriented Programming (AOP) paradigm, which provides abstractions for implementing autonomous and social software artifacts (agents).
However, the scope of EMAS entails more than AOP, in particular because EMAS is concerned with the holistic software engineering perspective and not only programming.
With the increase in prevalence of (somewhat) autonomous software systems in distributed information system landscapes~\cite{DelaPrieta2019}, it was initially a reasonable expectation that EMAS would gain attention from the software engineering mainstream.
However, EMAS approaches have not seen wide-spread adoption in practice, neither directly, nor as derivations that are implemented in industry-scale programming language ecosystems.

One focus area of EMAS is concerned with the design and development of methods and tools for AOP, examples of which are the Java-based Jade~\cite{Bellifemine2001} and JACK~\cite{Winikoff2005}, and the Belief-Desire-Intention (BDI)-based Jason~\cite{Bordini2007} frameworks\footnote{For an overview of some agent-oriented programming languages and frameworks, see (for example) Cardoso and Ferrando~\cite{cardoso2021review}}.
Indeed, the EMAS community has experimented with a broad range of abstractions for solving diverse problems using agents. However, no major success stories with regards to the establishment of industry-level tools, languages, or standards have been achieved; in their current state, the concepts and re-usable software libraries and frameworks provided by EMAS are still detached from the software engineering mainstream.
According to an EMAS community report~\cite{engineering-gsi-article-2019}, one of the key points of criticism of the state of affairs of EMAS is the lack of integration between EMAS and more widespread software engineering approaches.
Some recent works aim to address this issue, for example by integrating agent-oriented programming and modern \textquote{high-level} programming languages like JavaScript~\cite{bosse2016mobile,6722018,10.1007/978-3-030-51417-4_11}, and by providing resource-oriented abstractions to interact with autonomous agents and multi-agent systems~\cite{10.1007/978-3-319-91899-0_8}.
While these (and similar) tools help push the frontier of modern agent-oriented \emph{programming} towards practicality, no holistic agent-oriented perspectives on the complete software engineering life-cycle seem to exist.
An example of a deficiency that affects several steps of the life-cycle and that even the most mature AOP frameworks have regards the lack of facilities for testing goal-oriented software artifacts.
Seeing EMAS and AOP through the eyes of DevOps can potentially help identify new solution approaches to address such deficiencies.
\section{Integrating EMAS with DevOps}
\label{ag-dev-ops}

Let us highlight that the main objective of DevOps is not \emph{automation}, which could also be achieved with traditional, homogeneous team constellations, but rather \emph{autonomy} of teams within a software development organization, which is achieved by relying on automation technologies.
From the description provided in Subsection~\ref{background-devops}, one can see that DevOps is, in the way it is currently practiced, concerned with autonomy on three levels:

\begin{enumerate}
    \item On the \textbf{organization level}, DevOps facilitates team autonomy by avoiding the necessity of hand-overs between development, QA, and operations teams.
    \item On the \textbf{integration level}, DevOps allows for continuous integrations and deployments, avoiding manual steps and hand-overs in the pipeline from code check-in to system deployment.
    \item On the \textbf{operations level}, DevOps provides abstractions that allow operators to specify high-level infrastructure requirements and handles lower-level details like exact resource allocation and machine provisioning autonomously.
\end{enumerate}

In contrast, EMAS focuses on the autonomy of the agents, as \emph{software artifacts}, that a software engineering team or organization creates, \emph{i.e.}, it adds a fourth autonomy level to the three-level perspective DevOps provides.
Table~\ref{compare} shows an overview of the four levels and explains them by example.

\begin{figure}[ht]
	\centering
	\includegraphics[width=0.7\textwidth]{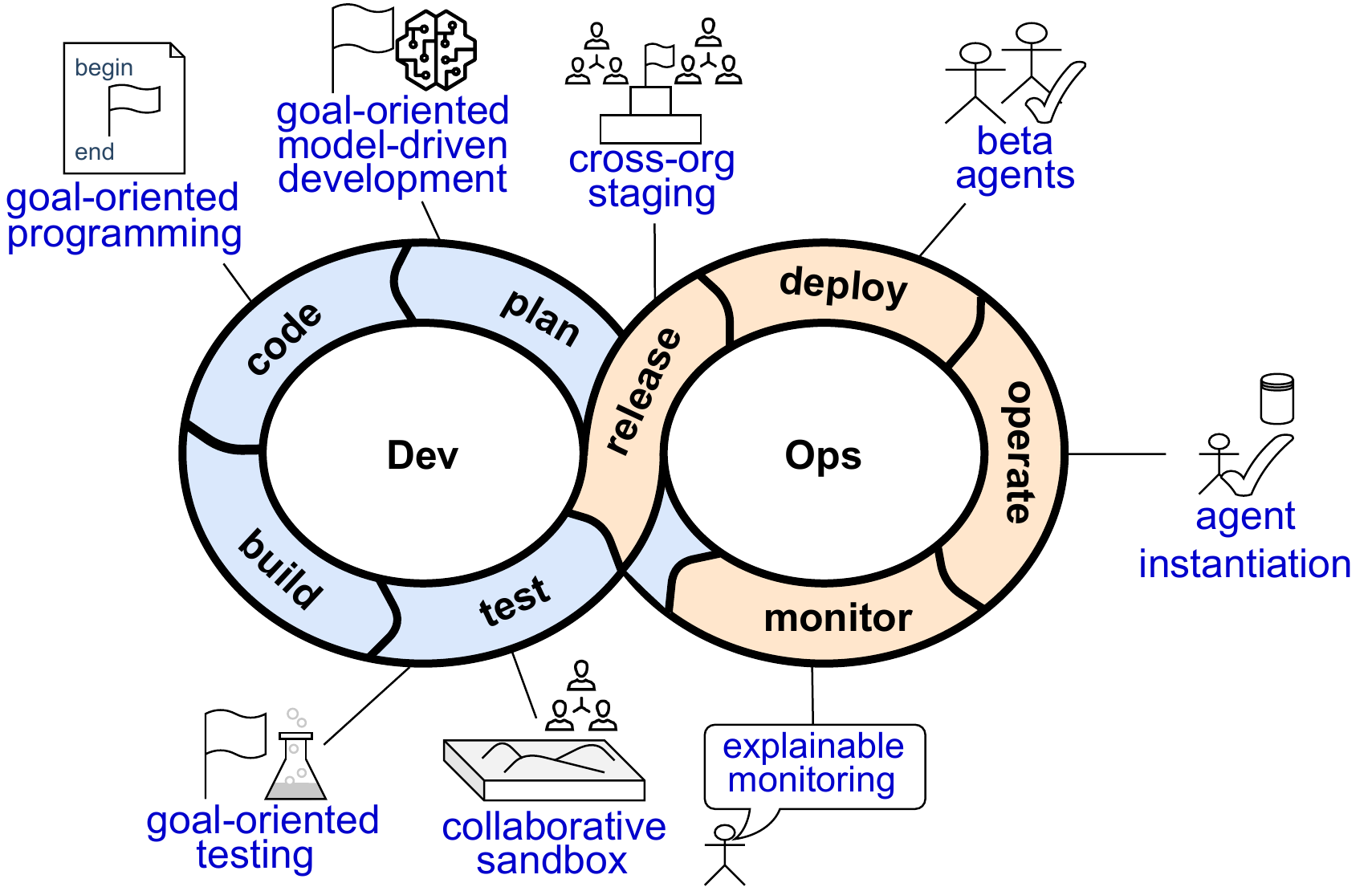}
	\caption{The DevOps life-cycle and agent orientation.}
	\label{fig:agentops}
\end{figure}

\begin{table*}
    \renewcommand{\arraystretch}{1.5}
    \caption{Autonomy levels, examples, and relevant approach.}
    \label{compare}
    \begin{tabular}{|p{2.5cm}| p{7.5cm}| p{1.75cm}|}
    \hline
    \textbf{Autonomy Level} & \textbf{Example} & \textbf{Approach} \\ 
    \hline
    Organization autonomy & 
    Avoid handovers: one team is in charge of all steps in the life-cycle &
    DevOps \\ 
    \hline
    Integration \phantom{eeeee} autonomy & 
    Avoid manual deployments and QA: run all tests before a merge and auto-deploy if all tests pass & 
    DevOps \\ 
    \hline
    Operations \phantom{eeeee} autonomy & 
    Avoid manual resource provisioning: auto-scale systems when load increases & 
    DevOps \\ \hline
    Artifact \phantom{eeeeeeee} autonomy & 
    Avoid manual low-level business decisions: approve (financial) transactions without humans interference & 
    EMAS \\
    \hline
    \end{tabular}
    
\end{table*}
At this point, it is worth highlighting that even when developing \textquote{traditional} software artifacts with little or no autonomy, it is widely acknowledged that total global supervision and coordination of all software design steps is practically not possible, even if the scope of the project is confined to a single organization. Hence, DevOps approaches try to integrate changes frequently in a controlled manner in order to discover unknown dependencies and unexpected behavior early on. The \emph{autonomy levels} (Table~\ref{compare}) allow teams of engineers to dynamically respond to challenges that arise and to minimize the effect these challenges have on the broader organization. When developing \emph{autonomous} software artifacts, one can expect that there will be even more problems that cannot be identified at design-time and hence the continuous integration approach requires even more attention; \emph{i.e.}, on the organizational level, the implementation of highly autonomous artifacts implies that the intensity of dependencies between teams that develop different sub-systems is not always apparent before these sub-systems are integrated.
These emerging dependencies then need to be managed on the integration and operations levels, for example to ensure that in case of the deployment of sub-systems that have \textquote{hidden} incompatibilities, communication failures do not lead to disastrous consequences.
Because of their dynamic nature, agents cannot be developed into mature software artifacts without exposing them to the environment they are supposed to act in~\cite{Winikoff2015}.
Regarding this distinguishing characteristic of agents, the integration of agent-orientation and developer operations can be considered a methodological response to this issue.
To allow for a gradual exposure of an agent to a progressively more realistic environment that increases the likelihood of catching critical errors early on, an agent-oriented variant of the DevOps life-cycle may require the following features:
\begin{itemize}
    \item \textbf{Goal-oriented test-driven development.} 
    The behavior of social and goal-oriented software artifacts like agents is typically complex and non-deterministic~\cite{Coelho2007}.
    Hence, the common testing levels (unit tests, functional tests, and integration tests) usually are not satisfactory to cover agents' possible behaviors. Some approaches have been proposed to address this issue~\cite{Coelho2007,Earle2019,Nguyen2009}, but a comprehensive solution has still not been devised~\cite{Winikoff2018}.
    \emph{Goal-oriented} tests can provide an extra test level that should be able to assess whether an agent's inference process from goals and beliefs to actions (and explanations of these actions) behaves as expected or not.

    \item \textbf{Sandbox for real-time collaboration.}
    Development teams can move agents that have passed static code analysis, unit tests, goal-oriented tests, and low-level integration tests (which may or may not be goal-oriented)\footnote{We assume that these tests can be executed relatively quickly when the developer logs a change, which is -- in the case of standard approaches to static code analysis (often called \emph{linting)}, unit tests, and some integration tests like micro-service handler tests -- a common capability of development tool-chains.} to a sand-boxed environment that allows for the collaborative development of agents and multi-agent systems in (near) real-time.
    This makes it easier for developers to consider their current development work in the context of other ongoing changes.
    Each sand-box features a fully-fledged multi-agent system, as well as version control and continuous integration support (automated testing and deployments).
    From a practical perspective, one can assume that the scope of a sandbox is restricted by organizational boundaries.
    For example, given a commercial enterprise A and a government organization B who both work on the same multi-agent system, it is safe to assume that the engineers of A cannot align in real-time with the engineers of B; a change made by organization A during development should not immediately (before verification and validation) affect the system organization B is developing against. 
   The EMAS community has presented an initial prototype addressing part of this issue~\cite{amaraldemoaamas}.
     
    \item \textbf{Cross-organizational staging system.}
    To ensure quality across organizational boundaries, stable versions of local agents, artifacts, and environment updates that have been developed and thoroughly tested in a sand-boxed environment can be deployed to cross-organizational staging systems.
    To these staging systems, organizations that depend on each other's work in a particularly critical manner (if not all organizations that contribute to the multi-agent system) have access and use it as a second-level testing environment; \emph{i.e.}, any run-time issue that may occur on the staging system does not effect system end-users. Still, errors are potentially more costly when they occur on the staging system and not in the sand-box, as their root-cause needs to be traced back -- in a more complex environment -- to a particular organization and then to a team.
    Cross-organizational staging systems can potentially make use of concepts and tools the EMAS community provides for managing multi-agent organizations (e.g., $\mathcal{M}$OISE$^+$~\cite{Hubner2007}).
    
    \item \textbf{Beta agents in production environments.}
    When the tests have passed in the cross-organizational staging environment, a step-wise production deployment can be executed.
    As a first step, agent instances can be exposed to \textquote{sense} the production environment, without being able to act upon it.
    Then, some agent instances can be fully deployed to the production environment, but at limited scale, analogously to the way beta-feature roll-outs are handled in many software-as-a-service environments (so-called \emph{canary deployments}~\cite{Bass2015}).
    Still, in contrast to typical canary deployments, which only affect a small portion of a system's users, beta agent deployments are potentially more critical because of the interconnectedness of multi-agent systems.
    Only if these beta agents pass all tests after extensive monitoring, the full update of the production environment is executed.
    This step reflects tests on real traffic scenarios used by the automotive industry~\cite{Huang2016,Stadler2016}\footnote{In Vehicle-in-the-loop (VEHIL) simulations, domain-specific concepts similar to the \textit{sandbox for real-time collaboration} and the \textit{cross-organizational staging system} are employed.}.
    
    \item \textbf{Explainable Monitoring}.
    Given the complex and non-deterministic behavior of multi-agent systems, it can be assumed that traditional monitoring facilities provide only limited utility.
    New ways of filtering and aggregating log entries for human or machine interpretation need to be devised.
    To address this issue, one can draw from an emerging body of works on explainable agents and multi-agent systems~\cite{10.5555/3306127.3331806}, and in particular from research that investigates the filtering of event data to generate human-digestible explanations~\cite{mualla2020human}.
\end{itemize}
Table~\ref{tab:features} list these features and provides an overview of how they relate to mainstream software engineering practices.
In the Figure~\ref{fig:agentops} we present a more comprehensive view of the development cycle of agents based on DevOps life-cycle.
Besides the mentioned features, the referred picture also illustrates the place for goal-oriented/agent-oriented model-driven development and programming tools, well-covered subjects of a range of studies produced by EMAS community (e.g., \cite{Bellifemine2001,Bordini2007,DeLoach2010,Uez2014,Winikoff2005}).

\begin{table}
\centering
    \renewcommand{\arraystretch}{1.5}
    \caption{Integrating AOP and DevOps: example features in comparison to existing practices.}
    \label{tab:features}
    \begin{tabular}{|p{3cm}| p{3cm}| p{3cm}|  p{3cm}|}
    \hline
    \textbf{Feature} & \textbf{Existing Practice} & \textbf{Similarities} & \textbf{Differences} \\ \hline
    Goal-oriented test-driven development & 
    Test-driven development (unit tests) & 
    Testing before/while implementing & 
    Higher declarative abstraction level at the intersection of unit and integration testing \\ \hline
    Sandbox for real-time collaboration & 
    Sandbox for exploratory development & 
    Rapid prototyping support & 
    Near-real-time interactive and collaborative programming support \\ \hline
    Cross-organizational staging system & 
    Traditional staging system & 
    Production-like environment & 
    Continuous deployments by different organizations \\ \hline
    Beta agents in production environments & 
    Beta-features in production environments & 
    Pilot beta-features in production environment & 
    Interaction between beta-agents and stable agents \\ \hline
    Explainable monitoring & 
    Operations monitoring systems & 
    Explanation/analysis of a running system & 
    System-centered versus agent-centered perspectives  \\ \hline
    \end{tabular}
\end{table}
Let us highlight that the list of features is primarily an initial starting point, and each feature comes with limitations and trade-offs that may only emerge in industrial application scenarios (and may be specific to a given domain, technology stack or DevOps-variant).
Consequently, it can make sense to consider a step-wise introduction of agent-oriented approaches to DevOps, focusing on the controlled assessment of a \emph{minimally viable} agent-oriented abstraction\footnote{In his \emph{Agent Programming Manifesto}, Logan calls for \emph{modular} approaches to AOP~\cite{logan2018agent}. We argue that the notion of a \emph{minimally viable abstraction} goes a step further, as it suggests a focus on one particular benefit AOP can bring to mainstream software engineering approaches such as DevOps, and hence a radical simplification that may deliberately disregard many aspects of AOP to minimize technology overhead and learning curve when introducing a single abstraction.}.
For example, autonomous software systems that are not developed using an academic EMAS approach can potentially still be evaluated by goal-oriented tests.
\section{Implications for EMAS}
Traditionally, EMAS is primarily concerned with the implementation of theoretical perspectives, such as belief-desire-intention reasoning-loops, that the artificial intelligence scientific literature provides on the design of autonomous agents and multi-agent systems.
In contrast, the approach outlined in this paper is pragmatically targeted at moving EMAS closer to modern industry practices for software development, and at identifying gaps in mainstream software development approaches and frameworks that EMAS can fill.
Hence, the approach depends on the exposure of EMAS and AOP works to the context of mainstream software development tools and pipelines, and in particular to the technology ecosystem that has risen to popularity alongside with DevOps.
First prototypes that work towards this goal by treating continuous integration, collaboration features, and distributed version control as first-class citizens in the context of agent-oriented programming exist~\cite{amaral2020,amaraldemoaamas}. 

Consequently, the whole technology ecosystem that makes up the DevOps tool-chains needs to be thoroughly analyzed, and methodologies and re-usable software frameworks (or framework extensions) for identifying and addressing the specific requirements for the DevOps-oriented management of goal-oriented, autonomous software artifacts need to be developed.
Logging, monitoring, and debugging facilities need to be devised that address the challenge of identifying anomalous behavior in a highly dynamic and heterogeneous environment, and facilitate the identification of software bugs that may be caused by intractable state and software version dependencies between autonomous software agents that are developed by different organizations.

Nevertheless, let us highlight that the integration of EMAS and DevOps cannot only draw from AOP research, but also apply other fundamental research on autonomous agents and multi-agent systems, for example by considering fundamental theoretical research on topics like belief revision~\cite{10.5555/1643031.1643098}, goal reasoning~\cite{aha2018goal}, or agreement technologies~\cite{10.5555/2431387}.
Still, EMAS and EMAS-related research that is of immediate relevance necessarily has a focus on technologies, software engineering processes and/or practical aspects of socio-technical systems. In contrast, research that primarily provides formal contributions would first need to be implemented as a generic and re-usable abstraction for a particular technology ecosystem, or be presented as a solution to a particular software engineering problem. In this context, the notion of a \emph{minimally viable abstractions} may -- again -- serve as a guiding design principle; \emph{e.g.}, when devising a new formal approach to belief revision, it may not be necessary to provide a holistic integration with a full-fledged MAS conceptual meta-model and technology like JaCaMo. Instead, a small library for managing belief revision could be implemented and presented in a way that enables re-usability in software stacks and tool-chains that do not necessarily include other agent-oriented concepts or technologies.

\section{Conclusion}
In this paper, we have proposed the integration of approaches to engineering multi-agent systems with the DevOps software engineering practice.
The integration expands the scope of the agent-oriented programming paradigm to cover the full life-cycle of modern software engineering, from initial specification via implementation and continuous integration to operation and monitoring.
Viewing EMAS from the perspective of modern software engineering approaches that cover the whole engineering life-cycle can facilitate the development of more practice-oriented perspectives on EMAS and AOP.
The integration of EMAS and DevOps can draw from the breadth and depth of research on agents and multi-agent systems, and motivate future work at the intersection of theory and practice, for example on goal-oriented testing and goal reasoning.

\subsubsection{Acknowledgments}
This work was partially supported by the Wallenberg AI, Autonomous Systems and Software Program (WASP) funded by the Knut and Alice Wallenberg Foundation and partially funded by Project AG-BR of Petrobras and by the program PrInt CAPES-UFSC ``Automação 4.0''.
%
%
%
\bibliographystyle{splncs04}
\bibliography{references}
\end{document}